# Investigating red packet fraud in android applications: insights from user reviews

Yu Cheng[1], Xiaofang Qi[1*] and Yanhui Li[2]

**Abstract**

With the popularization of smartphones, red packets have been widely used in mobile apps. However, the issues of fraud associated with them have also become increasingly prominent. As reported in user reviews from mobile app markets, many users have complained about experiencing red packet fraud and being persistently troubled by fraudulent red packets. To uncover this phenomenon, we conduct the first investigation into an extensive collection of user reviews on apps with red packets. In this paper, we first propose a novel automated approach, ReckDetector, for effectively identifying apps with red packets from app markets. We then collect over 360,000 real user reviews from 334 apps with red packets available on Google Play and three popular alternative Android app markets. We preprocess the user reviews to extract those related to red packets and fine-tune a pre-trained BERT model to identify negative reviews. Finally, based on semantic analysis, we have summarized six distinct categories of red packet fraud issues reported by users. Through our study, we found that red packet fraud is highly prevalent, significantly impacting user experience and damaging the reputation of apps. Moreover, red packets have been widely exploited by unscrupulous app developers as a deceptive incentive mechanism to entice users into completing their designated tasks, thereby maximizing their profits.

## Introduction

Mobile applications (apps) have become an essential part of our daily lives. In recent years, red packets, the traditional Chinese gifts of money enclosed in red envelopes, have been digitized in mobile apps (TechCrunch 2016). People can use apps such as WeChat (Wikipedia 2025c) and Alipay (Wikipedia 2025a) to send and receive red packets to each other. With the growing popularity of digital red packets, an increasing number of app developers distribute red packets containing cash or coupons within their apps to attract users to download and use the apps, which has led to the widespread presence of red packets across a wide variety of apps. However, as reported by the media (Paper 2024), some unscrupulous app developers exploit red packets as a deceptive incentive mechanism to defraud users for profit. For example, users are lured into completing a series of tasks, such as watching video advertisements (ads) or inviting new users to download the app, but ultimately fail to receive the corresponding red packet rewards as promised. From user reviews involving red packets on app markets, we have also found numerous complaints from users who report falling victim to red packet fraud and being troubled by fraudulent red packets. This phenomenon seriously affects the user experience, undermines users interests, and has a significant negative impact on the reputation of both the app itself and the app market. It is very necessary to expose the phenomenon of red packet fraud in the mobile app ecosystem.

Unfortunately, existing studies have not investigated the issues of red packet fraud that users frequently report

*Correspondence:
Xiaofang Qi
xfqi@seu.edu.cn
[1] School of Computer Science and Engineering, Southeast University, Nanjing 211189, China
[2] State Key Laboratory for Novel Software Technology, Nanjing University, Nanjing 210023, China

on app markets. The only study on red packet fraud focuses on the use of fake red packets as bait to spread malicious content (Cheng et al. 2025), such as malicious apps or malicious redirects. This study proposed an automated approach for detecting red packet fraud that propagates malicious content by analyzing the landing pages of red packets after they are clicked. However, this method cannot be utilized to investigate the fraudulent issues that users complain about in their reviews, which primarily concern the amounts of red packets and misleading promotional claims. As such, there is an urgent need for new methods to systematically explore these issues. In app markets, all users are free to post their reviews on the apps they use, including praise, complaints, and suggestions. These user reviews provide critical information for app developers to improve their apps, and as a result, user reviews are widely utilized in research focused on app development, evolution, and maintenance (Liu et al. 2023; Haggag et al. 2022; Li et al. 2020; Hu et al. 2021a; Le et al. 2023). To the best of our knowledge, no prior work in our research community has analyzed user reviews related to red packets. In this paper, we conduct the first empirical study and analysis of an extensive collection of user reviews on apps with red packets from popular Android app markets.

There are two key challenges in investigating the issues of red packet fraud from user reviews. The first challenge is how to collect user reviews related to red packets from the vast amount of user reviews in app markets. Since app markets host millions of mobile apps, most of which do not incorporate red packets, the review sections of most apps are unlikely to contain user reviews related to red packets. How to efficiently collect relevant user reviews poses a challenge. The second challenge is how to automatically identify red packet fraud issues from user reviews related to red packets. Given that a user review related to the red packet may discuss multiple features or aspects of an app rather than solely describing the red packet itself, it is necessary to filter out unrelated noise. Furthermore, accurately categorizing red packet fraud issues and summarizing the content described in each category presents a significant challenge.

In this paper, we design and implement an automated system consisting of three main steps to address the two challenges. We first propose a novel approach, ReckDetector, to effectively identify apps with red packets from app markets. Then, we employ web crawling techniques and relevant APIs to automatically collect real user reviews from apps with red packets on Google Play (Google 2025b) and three popular Chinese Android app markets, including Tencent Market (Tencent 2025), Huawei Market (Huawei 2025), and Xiaomi Market (Xiaomi 2025). Through keyword filtering and review segmentation, we extract the review content that only describes red packets (red packet reviews for short) from all user reviews. Subsequently, we fine-tune a pre-trained language model to identify negative reviews from red packet reviews. Finally, based on semantic analysis, these negative reviews about red packets are summarized into six distinct categories of red packet fraud issues reported by users.

The contributions of our work are as follows:

- We propose a novel automated approach, ReckDetector, for effectively identifying apps with red packets. We implement a prototype tool of ReckDetector, which is publicly available at https://github.com/AppFraud/ReckDetector. Experiments on the dataset of hundreds of real-world apps demonstrate that ReckDetector achieves higher performance than the state-of-the-art tool ReckDroid in identifying red packets.
- We conduct the first analysis of user reviews related to red packets by leveraging natural language processing and machine learning techniques, uncovering six categories of red packet fraud issues from real user reviews in app markets.
- Through our study, we found that red packet fraud is highly prevalent, significantly impacting user experience and damaging the reputation of apps. Moreover, red packets have been widely exploited by unscrupulous app developers as a deceptive incentive mechanism to entice users into completing their designated tasks, thereby maximizing their profits.

The rest of this paper is organized as follows. We first provide the background and motivation of our study in Section 2, and then detail our research methodology in Section 3. The results of our investigation are described in Section 4 and some discussion is provided in Section 5. Finally, Section 6 surveys the related work, and Section 7 concludes our paper.

## Background and motivation

### Red packet fraud

Red packet fraud is a broad and evolving concept that has emerged amid the growing prevalence of online scams. Given its dynamic characteristics and the continuous development of fraudulent tactics, there is currently no standardized or unified definition of red packet fraud. In this study, we adopt a working definition derived from empirical observations and common fraud patterns reported on popular platforms or media (TechinAsia 2016; Daily 2016; Paper 2020, 2024). Specifically, we refer to red packet fraud as deceptive activities that exploit digital red packets as bait or misleading incentive mechanisms to induce users to disclose personal information, click on phishing links, or complete designated tasks without the corresponding compensation that ultimately profit unscrupulous app developers.

### Research motivation

To increase app downloads and user engagement, some app developers promise to distribute red packets to users in their apps. However, users may feel deeply disappointed if the opened red packets are empty or the received amount is significantly lower than what was promised by the developers, as this constitutes a substantial deception. This is particularly true when users invest a considerable amount of time completing the required steps or tasks to receive the red packet, only to receive no reward in return.

User reviews are the simplest and most effective way to gather user feedback. As shown in Fig. 1, we list three examples of negative user reviews (one-star or two-star ratings) from three apps with red packets. In Fig. 1a and b, users complained that they received nothing when they opened the red packet. The user in Fig. 1c voiced a grievance that despite completing the required task (i.e., inviting new users) but the corresponding red packet reward was not received. It is noteworthy that the user reviews in Fig. 1a and c have garnered hundreds of likes, indicating that many other users have encountered similar situations. In fact, such user reviews are commonplace on app markets, yet they have received little attention from the research community. To uncover the phenomenon of red packet fraud, we have conducted an comprehensive investigation from the perspective of the users.

## Methodology

Our research methodology primarily consists of the following three steps, as illustrated in Fig. 2.

(1) Collection of apps with red packets. We propose an novel approach, ReckDetector, to automatically identify and collect apps with red packets from Google Play and three popular alternative Android app markets.
(2) Collection of red packet reviews. We employ web crawling techniques and relevant APIs to automatically collect user reviews for the apps with red packets. We select user reviews related to red packets from all user reviews through keyword filtering, and then segment them to extract the review content that only describes red packets.
(3) Red packet review analysis. Using machine learning methods, we identify negative red packet reviews and classify them into different categories based on their semantic information. We then leverage a state-of-the-art large language model (LLM) to generate concise summaries for each category.

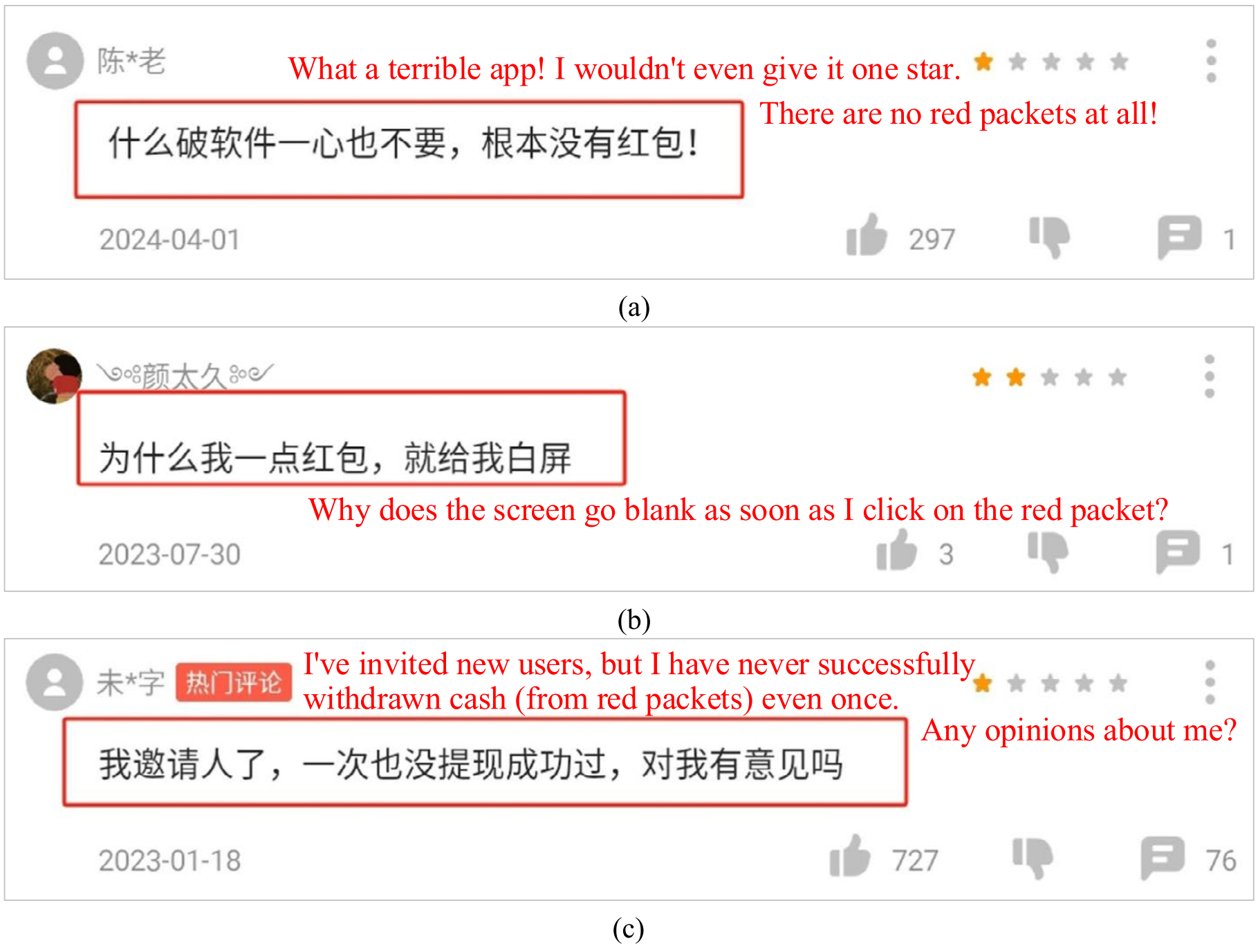

**Fig. 1** Examples of negative user reviews from three apps with red packets

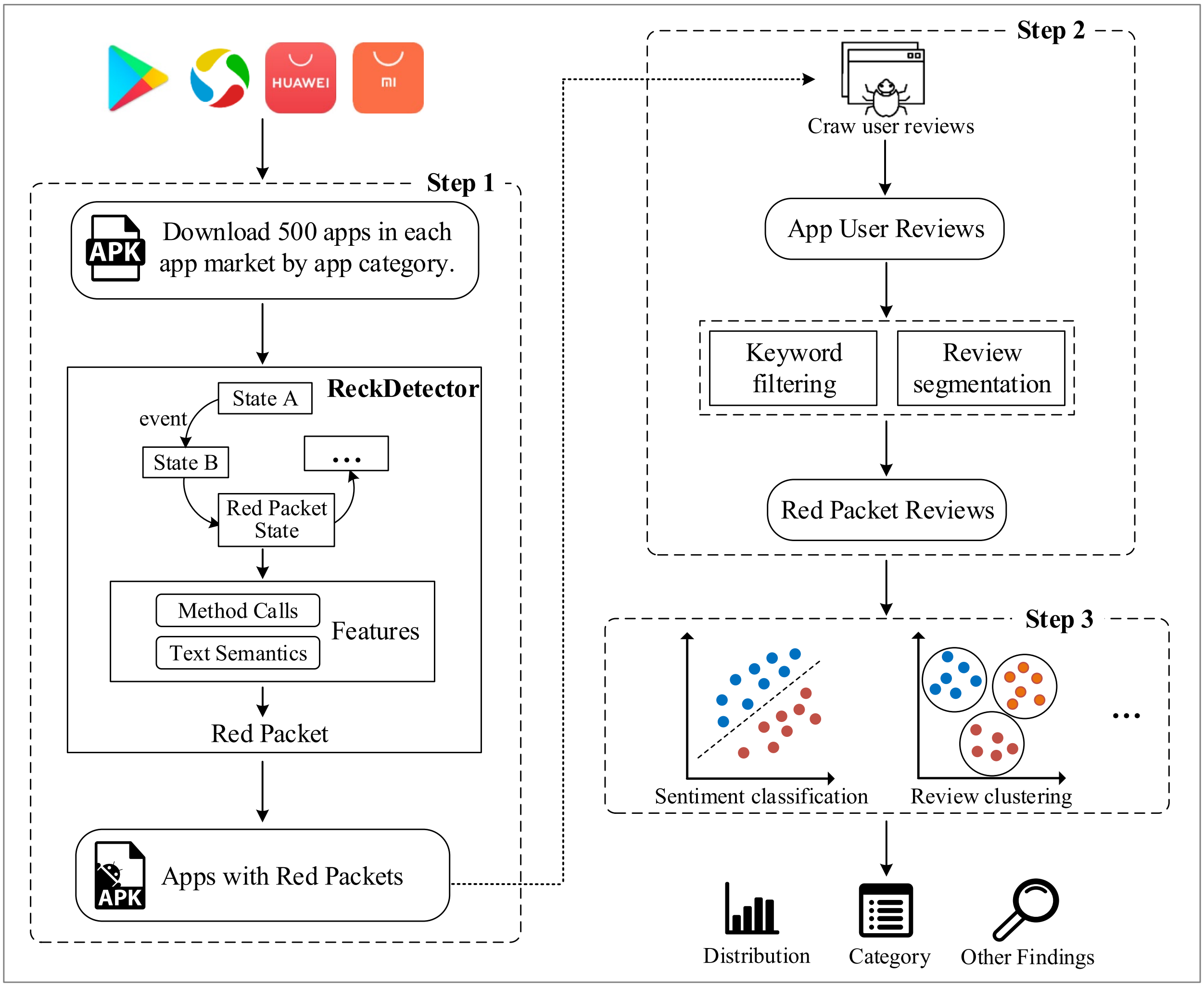

**Fig. 2** Overview of our research methodology

## Collection of apps with red packets

Since app markets do not have a dedicated app category for apps with red packets, we need to find a solution to collect such apps for our study. To this end, we propose a novel automated approach, ReckDetector, which is used to detect whether there are red packets in an app. Through ReckDetector, we can identify and collect apps with red packets from app markets.

### *ReckDetector*

Given that most apps that contain red packets do not have third-party libraries or APIs specifically related to red packets, it is not feasible to detect the presence of red packets through static analysis of the apps. Therefore, We adopt an automated exploration approach to dynamically run apps and trigger user interfaces (UIs) that may contain red packets, and then identify the pop-up red packets based on their features. Inspired by previous findings (Cheng et al. 2025) that most red packets appear in the first three layers of the app (mainly involving the main pages of the app and their subpages), our exploration is biased towards focusing on these pages. In this endeavor, we employ the red packet-first exploration strategy proposed in Cheng et al. (2025), as it has been proven to effectively trigger over 90% of red packets in apps.

It is well known that a red packet usually appears on the app's UI as a pop-up windows and contains some specific text, such as “Congrats! You've received a welcome red packet". We leverage these two key features to identify red packets, i.e., the methods invoked by the app when displaying pop-up windows and the semantic information of the text embedded in the pop-ups, as described in detail below:

*Method Calls:* At the Java code level of Android apps, we have identified and summarized four main strategies for implementing pop-up windows in the UI. These include system classes (e.g., Dialog), custom classes that extend system classes, third-party pop-up libraries (e.g., XPopup), and fully custom pop-up views. Table 1 presents some key attribute characteristics for different types of pop-up windows. They typically invoke different methods to be displayed on the app's UI. Therefore, we propose an API hooking approach to track all relevant methods used for displaying pop-up windows. Based on the hooking technique, we can dynamically detect whether there is a pop-up window in the current UI during app runtime.

**Table 1** Key attribute characteristics for different types of pop-up windows

| Type | Class name | Method name | Resource_id |
|---|---|---|---|
| System Class | Dialog, AlertDialog, PopupWindow, etc. | Show, showAsDropDown, showAtLocation, etc. | – |
| Custom Class | e.g., DialogNewGift, BubblePopupWindow, etc. | Show, showAsDropDown, showAtLocation, etc. | – |
| Third-party Library | XPopup, Material Dialogs, etc | OnShow, show, etc | – |
| Custom Pop-up View | FrameLayout, LinearLayout, etc | AddView, inflate, etc | Contain "pop" or "dialog" |

*Text Semantics:* The text embedded in red packets usually contains semantic information related to the red packet. Based on the analysis of the 550 different red packet images, Cheng et al. (2025) have summarized a representative set of 12 generic texts containing red packet semantics. Their study concluded that a semantic similarity score of 0.6 or higher with any of the 12 generic texts was established as the threshold for identifying red packet-related texts. Therefore, we utilize text semantics to determine whether a pop-up window is a red packet. By hooking the methods related to pop-up windows at the app layer, we can capture instances of pop-up windows and extract all text from these instances. We employ the Sentence-BERT model (Reimers and Gurevych 2019) to encode the text in a pop-up window into a text vector containing semantic information and separately calculate the cosine similarity scores with the 12 generic red packet texts. When the highest score among them is no less than 0.6, the pop-up window is considered a red packet.

### *App collection*

Since the Android platform has the largest market share worldwide (over 73.5% as of December 2024),[1] we decide to target only Android apps. We choose the official Android app market, Google Play, and three popular alternative Android app markets, including Tencent Market, Huawei Market, and Xiaomi Market. For each app market, we select 10 app categories spanning various domains, including shopping and payment, sports and fitness, video, reading, system tools, browsers, news, comprehensive services, leisure puzzle, and music. We then decide to randomly download 50 apps for each category, resulting in a total of 2,000 apps. Finally, we use ReckDetector to automatically identify apps with red packets from these apps.

### Collection of red packet reviews

We first obtain user reviews of the apps with red packets collected from the corresponding app markets. For Chinese app markets, we develop dedicated web crawlers to extract user reviews, while for the Google Play Store, we leverage an open-source library (Olano 2025) to access the review data. To extract user reviews related to red packet activities, we apply a keyword-based filtering approach to identify relevant reviews. Specifically, the keyword set includes not only explicit mentions such as "red packet", "cash", and "coins", but also implicit or indirect terms frequently associated with red packet-related issues, such as "fake", "rewards", "misleading", "trick", "withdraw", and "scam". Considering the linguistic differences between Chinese and English, the final keyword sets consist of 34 terms in English and 23 in Chinese.[2]

Since a user review may refer to several features or aspects of an app, rather than just the descriptions of red packets, we need to extract the part describing red packets from each review. To this end, we segment each review related to red packets into several short sentences based on punctuation and retain only those sentences containing the keywords. These sentences constitute the description of red packets in each user review, i.e., red packet reviews. After identifying all red packet reviews, we translate all non-English reviews into English using the Google API (Google 2025a) to ensure consistency for subsequent analysis and processing.

### Red packet review analysis

Since descriptions about red packet fraud primarily appear in negative user reviews, we focus exclusively on those negative reviews related to red packets. In each review, users have given a rating from one to five stars, where a rating of one or two stars usually indicates a negative user feedback. However, through manual inspection of the descriptions and ratings of some user reviews, we found that the two are not strictly consistent. For example, a negative feedback may occasionally be mistakenly rated three to five stars by a user. In addition, high-rating user reviews containing red packets may pertain to praise for other features within the app rather than the red packets themselves. Therefore, we cannot simply rely on user ratings to identify negative red packets reviews, as this may overlook some negative feedback regarding red packets. It is necessary to reclassify red packet reviews as either negative or positive. After obtaining negative red packet reviews, we will further investigate the various categories of red packet fraud issues reported in user reviews.

[1] https://gs.statcounter.com/os-market-share/mobile/worldwide

[2] https://github.com/AppFraud/ReckDetector/tree/main/datasets/keywords

#### *Sentiment classification*

To effectively identify negative red packet reviews, we adopt a deep learning-based approach. Specifically, we fine-tune a pre-trained BERT model (Devlin et al. 2019) to perform sentiment classification on review texts, classifying them as either negative or non-negative (including positive and neutral). To fine-tune the model, we manually label a subset of the collected red packet reviews to serve as the training set. We then employ the fine-tuned BERT model to identify negative reviews from all red packet reviews.

#### *Clustering analysis*

To uncover the phenomenon of red packet fraud from user reviews, we conduct a clustering analysis of negative red packet reviews based on textual semantics, aiming to investigate the different categories of red packet fraud issues reported in user reviews. We employ the Sentence-BERT model to encode all negative red packet reviews into text vectors, and then perform clustering on these text vectors through the K-Means algorithm. After determining the optimal number of clusters, we leverage the state-of-the-art LLM, GPT-4 (Petruzzellis et al. 2024), to extract the common semantics of red packet reviews contained in each cluster and generate concise summaries.

## Result analysis

In this paper, we mainly investigated the following three research questions (RQs):

- **RQ1:** *How many apps with red packets can our approach identify in app markets?*
- **RQ2:** *What is the distribution of user reviews related to red packet fraud and the associated apps across different app markets and app categories?*
- **RQ3:** *What categories of red packet fraud have been reported in user reviews?*

### RQ1: number of apps with red packets identified by our approach in app markets

*Motivation and Approach* RQ1 aims to investigate how many apps with red packets our approach can identify in app markets. As introduced in Section 3.1.2, we have collected 2,000 apps from four popular Android app markets, including Google Play, Tencent Market, Huawei Market, and Xiaomi Market. We employed ReckDetector to run these 2,000 apps on smartphones. Considering that some apps were unable to be properly accessed without user login, we manually logged into them before running these apps with ReckDetector.

*Results* Table 2 shows the distribution of apps with red packets identified from different app markets. Our approach identified a total of 334 apps with red packets, which account for 16.7% of all collected apps. The results indicate that apps with red packets are significantly prevalent, especially in major Chinese app markets, where their average proportion exceeds 20%. Specifically, among the four popular app markets examined, Tencent Market exhibits the highest proportion of apps with red packets at 24.8%, while Google Play has the lowest at 3.2%. This disparity is attributed to the fact that red packets originate from traditional Chinese culture and are primarily featured and used in apps from Chinese app markets, consequently resulting in a minimal presence of apps with red packet on Google Play.

### RQ2: distribution of user reviews related to red packet fraud and the associated apps across app markets and categories

*Motivation and Approach* RQ2 aims to investigate the distribution of user reviews related to red packet fraud and the associated apps across different app markets and app categories. We first crawled 361,580 user reviews from the 334 apps with red packets in the corresponding app markets. Through keyword filtering, a total of 54,763 user reviews containing red packets were obtained. Then we segmented each user review to extract the content that describes red packets, resulting in 54,763 red packet reviews. Given that negative red packet reviews typically describe instances of red packet fraud, we fine-tuned a pre-trained BERT model to identify such negative reviews. Specifically, we randomly sampled thousands of reviews from a corpus of over 50,000 red packet reviews and manually labeled 2,500 negative reviews and 2,500 non-negative reviews to construct our dataset. The annotation was guided by a set of predefined criteria to distinguish between negative and non-negative reviews:

- *Negative reviews* were defined as those expressing dissatisfaction, complaints, distrust, disappointment, or frustration regarding the red packet features, promotions, rewards, or withdrawal process. Common indicators include negative sentiment words (e.g., scam, fake, less, cannot withdraw, useless), sarcastic remarks, or explicit grievances about red packet activities.
- *Non-negative reviews* included neutral, positive, or ambiguous statements that did not explicitly convey dissatisfaction. These may express praise, curiosity, questions, or neutral descriptions without evaluative sentiment.

**Table 2** Distribution of apps with red packets identified from different app markets

| Market | # Apps | # Apps with red packets | Pct |
|---|---|---|---|
| Tencent | 500 | 124 | 24.8% |
| Huawei | 500 | 101 | 20.2% |
| Xiaomi | 500 | 93 | 18.6% |
| Google Play | 500 | 16 | 3.2% |
| Total | 2,000 | 334 | 16.7% |

To reduce labeling subjectivity, two independent annotators labeled all samples, and disagreements were resolved through discussion. This process helped ensure annotation consistency and label quality for subsequent supervised learning tasks.

*Results* We applied the fine-tuned BERT model to the remaining 49,763 red packet reviews, resulting in a total of 18,205 negative reviews, which includes the initially labeled 2,500 negative reviews. Table 3 shows the distribution of user reviews related to red packet fraud and the associated apps across different app markets. The fourth column represents the number of low ratings (1-star or 2-star) given by users for reviews containing red packets. The fifth column represents the number of predicted negative red packet reviews, which refer to user reviews related to red packet fraud. The sixth column indicates the proportion of user reviews related to red packet fraud among red packet reviews collected from each app market. The eighth column indicates the number of apps with user reviews mentioning red packet fraud. The ninth column represents the proportion of apps with user reviews mentioning red packet fraud among apps with red packet collected from each app market. The results show that, on average, 33.2% of user reviews mentioning red packets in app markets are related to complaints about red packet fraud. This suggests the widespread presence of red packet fraud among users. Specifically, among the four app markets analyzed, Google Play Store has the highest proportion of user complaints, accounting for 55.0%. Among the three Chinese app markets, Xiaomi Market exhibits the highest complaint rate, reaching 34.0%. At the app level, an average of 70.7% of apps with red packets from four app markets have been reported by users for red packet fraud issues. The proportion of apps receiving negative feedback varies across platforms, with the lowest observed in Tencent Market (59.7%) and the highest in Huawei Market (85.1%). These findings indicate that issues related to red packets are widespread and not limited to a specific market.

It is noteworthy that, as shown in Table 3, out of 54,763 user reviews related to red packets, 15,196 (27.7%) are given low ratings by users. This indicates, to some extent, that red packet fraud impacts user experience, resulting in substantial dissatisfaction among users. Moreover, such a high rate of low ratings may severely damage the reputation of the apps themselves and even undermine the credibility of the app markets in which they are hosted.

Table 4 shows the distribution of user reviews related to red packet fraud and the associated apps across different app categories. Given that the number of red packet reviews under the system tools category is extremely small, we focus only on the results from the other nine

**Table 3** Distribution of user reviews related to red packet fraud and the associated apps across different app markets

| Market | User review level | | | | | App level | | |
|---|---|---|---|---|---|---|---|---|
| | #User reviews | #Red packet reviews | #Ratings (≤ 2★) | #Negative reviews | Pct. | #Apps with red packets | #Apps with Neg. reviews | Pct. |
| Tencent | 108,918 | 13,670 | 3,377 | 4,550 | 33.3% | 124 | 74 | 59.7% |
| Huawei | 58,432 | 18,040 | 4,050 | 5,721 | 31.7% | 101 | 86 | 85.1% |
| Xiaomi | 158,632 | 22,624 | 7,480 | 7,698 | 34.0% | 93 | 64 | 68.8% |
| Google Play | 35,598 | 429 | 289 | 236 | 55.0% | 16 | 12 | 75.0% |
| Total | 361,580 | 54,763 | 15,196 | 18,205 | 33.2% | 334 | 236 | 70.7% |

**Table 4** Distribution of user reviews related to red packet fraud and the associated apps across different app categories

| Category | User review level | | | App level | | |
|---|---|---|---|---|---|---|
| | **# Red packet reviews** | **# Negative reviews** | **Pct.** | **# Apps with red packets** | **# Apps with Neg. reviews** | **Pct.** |
| shopping & payment | 7,271 | 4,477 | 61.6% | 65 | 46 | 70.8% |
| sports & fitness | 320 | 115 | 35.9% | 31 | 11 | 35.5% |
| comprehensive services | 7,891 | 864 | 10.9% | 45 | 30 | 66.7% |
| leisure puzzle | 5,321 | 1,387 | 26.1% | 43 | 35 | 81.4% |
| video | 18,339 | 5,266 | 28.7% | 55 | 41 | 74.5% |
| reading | 6,582 | 2,548 | 38.7% | 31 | 23 | 74.2% |
| news | 2,669 | 1,060 | 39.7% | 28 | 22 | 78.6% |
| browsers | 1,967 | 1,254 | 63.8% | 16 | 13 | 81.3% |
| music | 4,369 | 1,228 | 28.1% | 13 | 13 | 100.0% |
| system tools | 34 | 6 | 17.6% | 7 | 2 | 28.6% |
| Total | 54,763 | 18,205 | 33.2% | 334 | 236 | 70.7% |

Below is a group of user reviews related to red packets. Please summarize the main issue that users are collectively complaining about. Generate one concise sentence that captures the shared concerns expressed in these reviews. Avoid repeating specific details or examples—instead, focus on abstracting a high-level theme.

User reviews:

[Insert all the reviews from this category here, separated by line breaks.]

**Fig. 3** Prompt template designed for GPT-4

Below is a group of user reviews related to red packets. Please summarize the main issue that users are collectively complaining about. Generate one concise sentence that captures the shared concerns expressed in these reviews. Avoid repeating specific details or examples—instead, focus on abstracting a high-level theme.

User reviews:

[The newcomer benefits are a scam.
Fake promotions are everywhere.
This is clearly false advertising.
Fake marketing pretending to give out money.
These money-making activities are all fake.
There's a lot of false advertising inside.
... ]

**Output:** False advertising surrounding red packet campaigns is prevalent.

**Fig. 4** Prompt-response example for summarizing user reviews

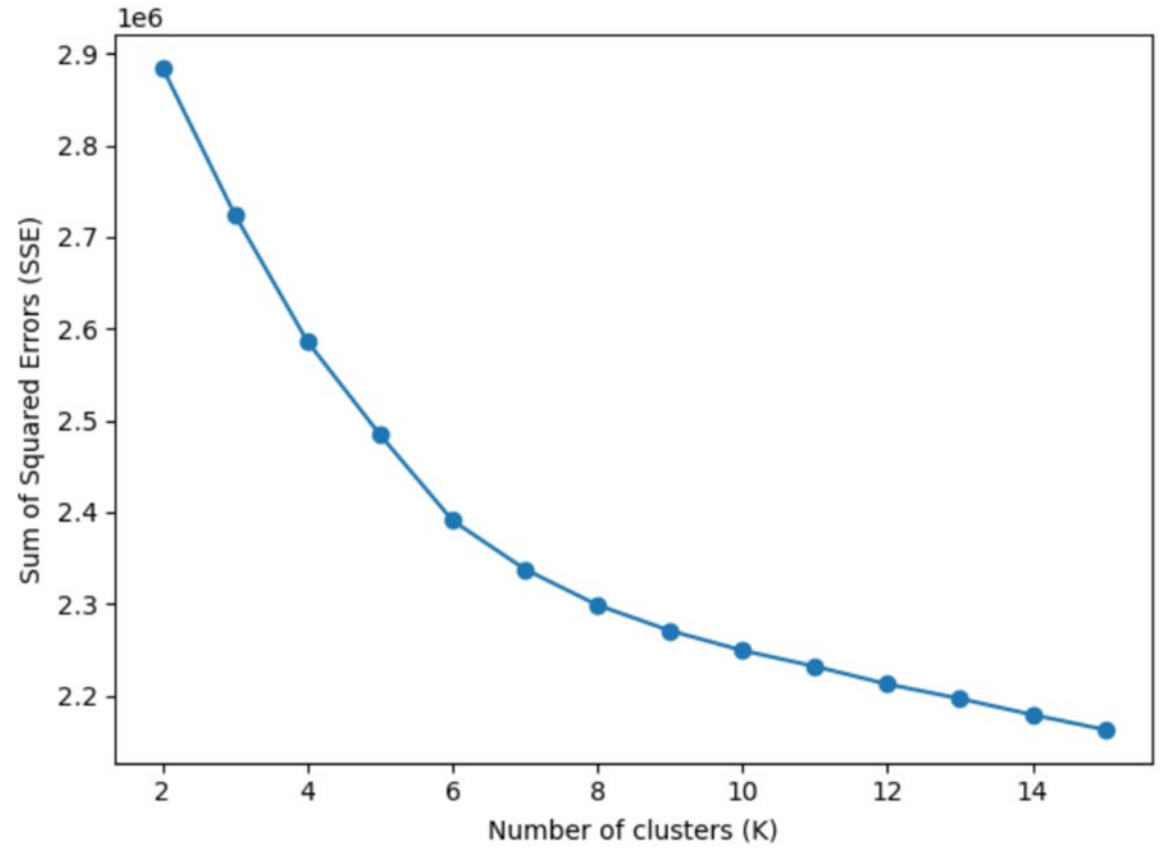

**Fig. 5** Elbow Method plot for determining the optimal number of clusters

app categories. The results indicate that the proportion of user reviews complaining red packet fraud under the browsers category is the highest (63.8%), followed by the shopping & payment (61.6%) and news (39.7%), reading (38.7%) categories. Developers of apps under these categories tend to release red packet tasks, such as viewing ads, browsing product pages, browsing news, or reading novels, to entice users to spend significant time on their apps, thereby increasing their revenue. However, in most cases, app developers fail to fulfill their promise of red packet rewards to users, leading to significant user dissatisfaction. In addition, we have found that music apps (100%) account for the highest proportion of apps reported by users for red packet fraud, followed by leisure puzzle (81.4%), browsers (81.3%) and news (78.6%) apps. In contrast, sports & fitness apps have the lowest proportions at 35.5%. These results suggest that user dissatisfaction with red packets is more prominent in apps that primarily focus on entertainment or content consumption, possibly due to the perceived misalignment between user expectations and actual reward experiences.

## RQ3: categories of red packet fraud reported in user reviews

*Motivation and Approach* RQ3 aims to investigate the categories of red packet fraud reported in user reviews. To this end, we classified reviews related to red packet fraud based on textual semantics. We employed the Sentence-BERT model to separately encode 18,205 negative red packet reviews into vector representations and then applied the K-Means algorithm to cluster these vectors. For each cluster, we leveraged GPT-4 to extract the common semantics of all user reviews contained in the cluster and summarize them into a concise sentence. To extract high-level semantic topics from each cluster of user complaints, we designed specific prompts for GPT-4 to summarize the core issue expressed in each group of reviews. Figure 3 presents the prompt template designed for GPT-4. Specifically, the prompt instructed the model to read a group of user reviews (within a single cluster) and produce one abstract sentence summarizing the central issue reflected in the reviews. The model was explicitly asked to avoid mentioning specific user examples or details, and instead to focus on abstracting a shared concern across the reviews. Figure 4 provides a prompt-response example for summarizing user reviews. In this example, we input the relevant review texts into a predefined prompt template, and GPT-4 generated a concise topic summary as the output.

**Table 5** Different categories of red packet fraud reported in user reviews

| No | Summary of reviews | # Reviews | Pct |
|---|---|---|---|
| 1 | Users are unable to receive any red packet rewards | 6,393 | 35.1% |
| 2 | Red packets contain an insufficient amount of gold coins | 3,220 | 17.7% |
| 3 | Red packets are difficult to cash out | 3,093 | 17.0% |
| 4 | It's hard to make any money from these red packet tasks. | 2,769 | 15.2% |
| 5 | The red packet promotions are full of tricks. | 1,426 | 7.8% |
| 6 | False advertising surrounding red packet campaigns is prevalent. | 1,304 | 7.2% |

*Results* Figure 5 represents a line plot obtained using the Elbow Method (Kodinariya and Makwana 2013), which is a commonly used technique for determining the optimal number of clusters in clustering algorithms. The x-axis indicates the number of clusters ($K$) into which the data is partitioned. The y-axis represents the sum of squared errors (SSE), which measures the total squared distance between each data point and the centroid of its assigned cluster. Lower SSE values indicate that the data points are closer to their respective centroids, suggesting better clustering. The "elbow" of the curve, where the rate of decrease sharply changes, is typically chosen as the optimal $K$ value. From Fig. 5, the elbow appears to be at $K = 6$, as the SSE reduction rate slows down significantly after this point.[3] Therefore, user reviews related to red packet fraud can be classified into six categories.

By inputting the red packet reviews from each cluster into GPT-4 using the designed prompt template, we obtained six concise topic summaries, each representing a distinct category of red packet fraud issues, as shown in Table 5. These reflect common complaints about red packets, including being unable to receive any red packet rewards, receiving insufficient amounts, difficulties in cashing out, low earnings from associated tasks, the prevalence of red packet tricks, and excessive fake advertising.

(1) No Red Packet Rewards. The most frequently reported issue (35.1%) is that users were unable to receive any red packet rewards. This issue may stem from fraudulent promotions, technical glitches, or intentional withholding of rewards by the app platform. Users may experience frustration and distrust, leading to an increase in complaints, negative reviews, and potential legal consequences.
(2) Insufficient Amount. The second most common complaint (17.7%) is that users received insufficient rewards when they opened red packets. Platforms attract users through exaggerated promotional strategies, but the red packets they provide often contain minimal monetary value, falling short of user expectations. This discrepancy between promised and actual rewards leads to user dissatisfaction, as it creates a sense of deception among users.
(3) Difficult to Cash Out. A significant portion of users (17.0%) have reported that it is difficult to cash out from red packets. This issue often arises from high withdrawal thresholds, complicated procedures, or hidden restrictions imposed by the app platform, making it challenging for users to access their funds. It undermines user trust, reduces engagement, and may result in damage to the app's reputation.
(4) Hard to Make Money. A substantial number of users (15.2%) expressed that it was hard to make any money from red packet tasks. They are required to complete designate tasks to receive red packets, but most tasks are unreasonably hard or even impossible to finish. Platforms tend to design overly complex or misleading red packet tasks, such as requir-ing numerous referrals or fabricating task chain traps,[4] to entice users to spend significant time and effort on their apps while simultaneously preventing them from obtaining rewards. Users may feel exploited or misled, resulting in negative feedback, reduced participation in future campaigns, and ethical concerns about app platform practices.

[3] We have used an automated method, KneeLocator (Satopaa et al. 2011), to confirm the elbow point.

(5) Full of Tricks. A notable subset of reviews (7.8%) described red packet promotions as being full of tricks. These tricks often involve unclear rules, hidden conditions, or manipulative task designs that mislead users. Such deceptive tactics diminish perceived fairness and contribute to a broader perception of the platform as untrustworthy.

(6) Pervasive False Advertising. A smaller but still meaningful group of users (7.2%) complained about pervasive false advertising in red packet campaigns. These reviews highlighted discrepancies between promotional claims and actual user experiences, including overstated rewards and misleading withdrawal conditions. Such practices erode user trust and may lead to accusations of dishonest practices and regulatory scrutiny.

Overall, these issues reflects primary user complaints regarding red packet fraud, suggesting that red packets have been widely adopted as a deceptive incentive mechanism to induce users to profit app developers. They seriously impact user experience, undermine users interests, and may lead to regulatory interventions or reputational damage if not addressed effectively.

## Discussion

In this section, we first evaluate the effectiveness of ReckDetector in identifying apps with red packets and our classification approach in identifying negative red packet reviews. Then we further characterize red packet fraud by analyzing high-frequency words in negative red packet reviews. To ensure the reliability of our findings, we include manual validation of GPT-generated summaries through random sampling and keyword-based checks, as well as manual verification of red packet fraud cases reported in user reviews. Additionally, We submit the apps reported for red packet fraud to VirusTotal for detection. Finally, we discuss the limitations of our study.

[4] Task chain trap: Each completed task unlocks a new one with higher rewards but escalating difficulty, ultimately making the final task nearly impossible to complete, thereby preventing users from earning any rewards.

**Table 6** Results of different methods for identifying red packets

| Method | Precision | Recall | F1-score |
|---|---|---|---|
| ReckDroid | 98.0% | 93.3% | 95.6% |
| ReckDetector | 98.5% | 98.1% | 98.3% |

### Effectiveness of ReckDetector in identifying apps with red packets

To verify whether ReckDetector can effectively identify apps with red packets, we evaluate it using the dataset provided in Cheng et al. (2025), including 144 apps with red packets and 250 apps without red packets. In addition, to demonstrate the performance and advantages of our approach, we conduct a comparison with ReckDroid in identifying red packets.

Table 6 presents a performance comparison between ReckDetector and ReckDroid in identifying red packets. The precision, recall and F1-score of ReckDetector are 98.5%, 98.1% and 98.3%, respectively, all of which surpass those of ReckDroid. In particular, ReckDetector significantly outperforms ReckDroid in terms of recall. As long as an app is found to contain at least one red packet, it is considered an app with red packets. Therefore, ReckDetector can identify apps with red packets effectively.

We further analyze the underlying reasons for this performance difference. ReckDroid (Cheng et al. 2025) uses three key features—UI widgets, color, and text—to identify red packets. In the evaluation experiment, ReckDroid generated 18 false negatives in identifying red packets. Of these, 12 resulted from UI Automator's failure to capture accurate UI view hierarchies for these apps, preventing the detection of pop-up views based on UI widget features. The remaining six were caused by the uncommon color of the red packets. Our approach identifies pop-up windows by dynamically tracking the methods called to display them, thereby avoiding false negatives caused by inaccurate UI view hierarchies. Moreover, we do not incorporate the color features of red packets into our detection criteria. Consequently, ReckDetector demonstrates a significantly lower rate of false negatives than ReckDroid.

### Effectiveness of our classification approach in identifying negative red packet reviews

To validate the effectiveness of our classification approach, we fine-tuned a pre-trained BERT model to identify negative red packet reviews and compared its performance against several representative machine learning classification models, including SVM, Random Forest, Naive Bayes, and Logistic Regression. As introduced in Section 4.2, we have manually collected and labeled 5,000 red packet reviews, including 2,500

**Table 7** Results of different methods for identifying negative red packet reviews

| Method | Precision | Recall | F1-score |
|---|---|---|---|
| SVM | 93.8% | 94.6% | 94.2% |
| Random Forest | 93.6% | 96.8% | 95.2% |
| Naive Bayes | 92.7% | 91.4% | 92.0% |
| Logistic Regression | 93.5% | 95.6% | 94.6% |
| BERT | **94.6%** | **97.6%** | **96.1%** |

The bold values in the table represent the highest performance obtained across all methods for each corresponding evaluation metric negative reviews and 2,500 non-negative reviews. For traditional machine learning classifiers, we employed the Sentence-BERT model to extract the semantic features of red packet reviews and generate high-dimensional vector representations. From these 5,000 labeled reviews, we randomly selected 2,000 negative reviews and 2,000 non-negative reviews to form a training set for training or fine-tuning these models. Then, we put the remaining 1,000 reviews into a testing set and applied the trained or fine-tuned models to it. As shown in Table 7, the BERT-based model achieved the highest scores across all evaluation metrics, with a precision of 94.6%, a recall of 97.6%, and an F1-score of 96.1%. These results clearly surpass those of traditional models, whose F1-scores ranged from 92.0% to 95.2%. Notably, BERT demonstrated a substantially higher recall, which is particularly important for minimizing false negatives when detecting user dissatisfaction. The superior performance of BERT confirms the advantage of leveraging deep contextualized language representations for sentiment classification in real-world, user-generated content. These findings demonstrate that our BERT-based classification approach is quite effective and reliable for identifying negative reviews from all red packet reviews.

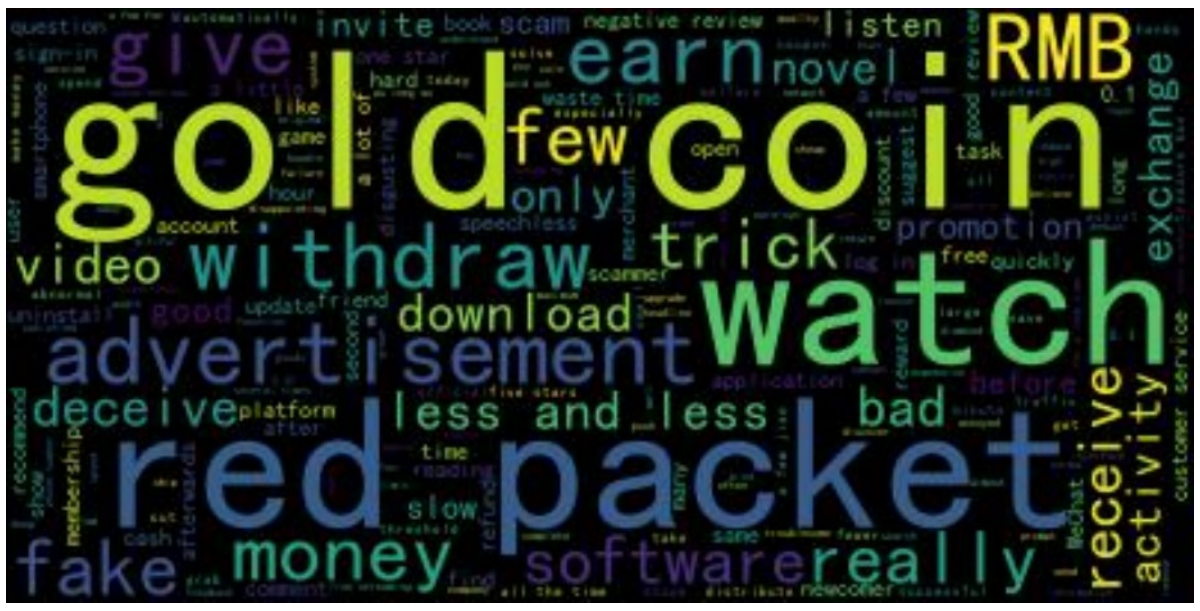

**Fig. 6** Word cloud from negative reviews about red packets

## Hot words in negative user reviews

To characterize red packet fraud, we further analyzed hot words in negative reviews related to red packets. We performed word segmentation for the 18,205 negative red packet reviews and extracted 289 high-frequency words, with the highest frequency being 6,952 times and the lowest being 50 times. Figure 6 gives the word cloud generated from these high-frequency words. In the word cloud, the bigger and bolder the word appears, the more often it's mentioned within user reviews. It can be seen that apart from "red packet" and "gold coin", words or phrases such as "few", "less and less", "trick", "fake", and "deceive" also appear frequently in user reviews. This phenomenon reflects widespread user complaints about red packets, revealing the fact that users are deeply plagued by fraudulent red packets. It also further indicates that negative reviews related to red packets often signify underlying issues of red packet fraud. In addition, words such as "watch", "advertisement", "video", "download", "withdraw", "invite" and "activity" are also often mentioned by users. This indicates that users usually need to complete specific tasks before receiving red packet rewards, such as watching video ads or inviting new users. These task often result in a poor user experience, especially when users spend a significant amount of time completing them but receive no corresponding red packet rewards. Overall, red packets have been widely exploited by unscrupulous app developers as a deceptive incentive mechanism to entice users into completing their designated tasks, thereby maximizing their profits.

## Manual validation of GPT-generated summaries

To ensure that the GPT-generated summary for each cluster accurately reflects the content of user reviews within that cluster, we conducted a two-step manual validation process.

### *Random sampling of user reviews*

For each of the six clusters, we randomly sampled 100 user reviews, resulting in a total of 600 reviews. We manually reviewed each sampled review and determined whether its content was consistent with the topic summary generated by GPT-4. To reduce subjective bias, two independent annotators were assigned to review all sampled reviews. Only those reviews for which both annotators reached agreement were retained as verified. In cases where their judgments differed, the two annotators discussed the disagreement and reached a consensus through joint deliberation. Table 8 presents the results of manual validation for the topic summaries generated by GPT-4. The column "# Consistent" indicates the number of reviews that both annotators agreed were consistent with the GPT summary. The column "Pct." shows the percentage of agreement for each cluster. Across all six clusters, consistency rates range from 89% to 100%, with an average agreement of approximately 95%. These results strongly support the accuracy and reliability of the GPT-generated topic summaries in capturing the core themes and concerns expressed in user reviews.

**Table 8** Manual validation results of GPT-generated summaries and the top 5 keywords of user reviews in each cluster

| No | # Samples | # Consistent | Pct | Keywords (Top 5) |
|---|---|---|---|---|
| 1 | 100 | 89 | 89% | gold coin, red packet, receive, reward, cash |
| 2 | 100 | 98 | 98% | gold coin, less and less, few coin, red packet, not enough |
| 3 | 100 | 95 | 95% | withdraw, cash out, gold coin, RMB, threshold |
| 4 | 100 | 91 | 91% | make money, earn, gold coin, ad, task |
| 5 | 100 | 100 | 100% | trick, trap, app, advertising, promotion |
| 6 | 100 | 96 | 96% | fake, promotion, false advertis-ing,<br>activity, scam |

#### *Keyword-based validation*

In addition to manual sampling, we extracted top keywords from each cluster using the keyword extraction technique TF-IDF (Wikipedia 2025b). These keywords were then compared against the GPT-summarized topic to check for semantic alignment and consistency. The last column in Table 8 lists the top five keywords extracted from each cluster of user reviews, representing the most prominent terms in each cluster. When compared with the GPT-generated topic summaries, there is a strong semantic alignment between the keywords and the summaries. For instance:

- *Cluster 1:* Keywords like "receive," "reward," and "red packet" directly support the summary "Users are unable to receive any red packet rewards."
- *Cluster 2:* Keywords such as "less and less," "few coin," and "not enough" match the summary "Red packets contain an insufficient amount of gold coins."
- *Cluster 3:* Keywords like "withdraw," "cash out," and "threshold" are consistent with "Red packets are difficult to cash out."
- *Cluster 4:* Terms such as "make money," "earn," and "task" reinforce "It's hard to make any money from these red packet tasks."
- *Cluster 5:* Words like "trick," "trap," and "advertising" support "The red packet promotions are full of tricks."
- *Cluster 6:* Keywords including "fake," "false advertising," and "scam" clearly validate the summary "False advertising surrounding red packet campaigns is prevalent."

This strong consistency between extracted keywords and the GPT-4 summaries provides additional empirical evidence that the summaries are not only accurate but also well-grounded in the most frequent user concerns expressed in each cluster.

### Manual verification of red packet fraud

To verify the authenticity of the red packet fraud reported in user reviews, we conducted a manual verification process. Table 9 presents the number of apps from each of the six categories of red packet fraud summarized in RQ3. It can be found that some apps contain more than one category of red packet fraud. We randomly selected 5 apps from each category of red packet fraud, resulting in a total of 30 apps. Since the occurrence of red packets in apps may be affected by factors such as the date, we manually ran these 30 apps three times over a span of one week, and then analyzed the behavior of receiving red packets and the corresponding amounts. Through careful manual verification, we have confirmed the presence of the six categories of red packet fraud within the corresponding 30 apps. This indicates that the red packet fraud

**Table 9** Number of corresponding apps under different categories of red packet fraud

| Category no | # Reviews | # Apps involved |
|---|---|---|
| 1 | 6,393 | 128 |
| 2 | 3,220 | 64 |
| 3 | 3,093 | 107 |
| 4 | 2,769 | 104 |
| 5 | 1,426 | 58 |
| 6 | 1,304 | 54 |
| Total | 18,205 | 236 |

reported by users in their reviews is generally authentic and reliable.

#### Detection results of VirusTotal

To further understand apps reported by users for red packet fraud, we employed VirusTotal (Virustotal 2025) to detect them. As shown in Table 3, we have collected 236 apps with negative red packet reviews. From these apps, we selected the top 10 apps with the highest number of negative red packet reviews and the top 10 apps with the highest proportion of negative red packet reviews from each app markets. After removing duplicates, a total of 52 different apps were obtained. We submited these 52 apps to VirusTotal for maliciousness detection. Figure 7 presents the VirusTotal detection results, showing that 25 apps (48.1%) were not labeled as malware by any anti-virus engine. This indicates that most anti-virus engines fail to identify malicious or suspicious behaviors in apps reported for red packet fraud. Given that some anti-virus engines on VirusTotal may not always report reliable results, we consider an app to be malicious only if at least three anti-virus engines flag it as malicious, based on previous works (Rastogi et al. 2016; Liu et al. 2020). However, only 11 out of the 52 apps (21.2%) were flagged by at least three anti-virus engines. This result suggests that the apps with red packet fraud cannot be sufficiently identified by existing anti-virus engines.

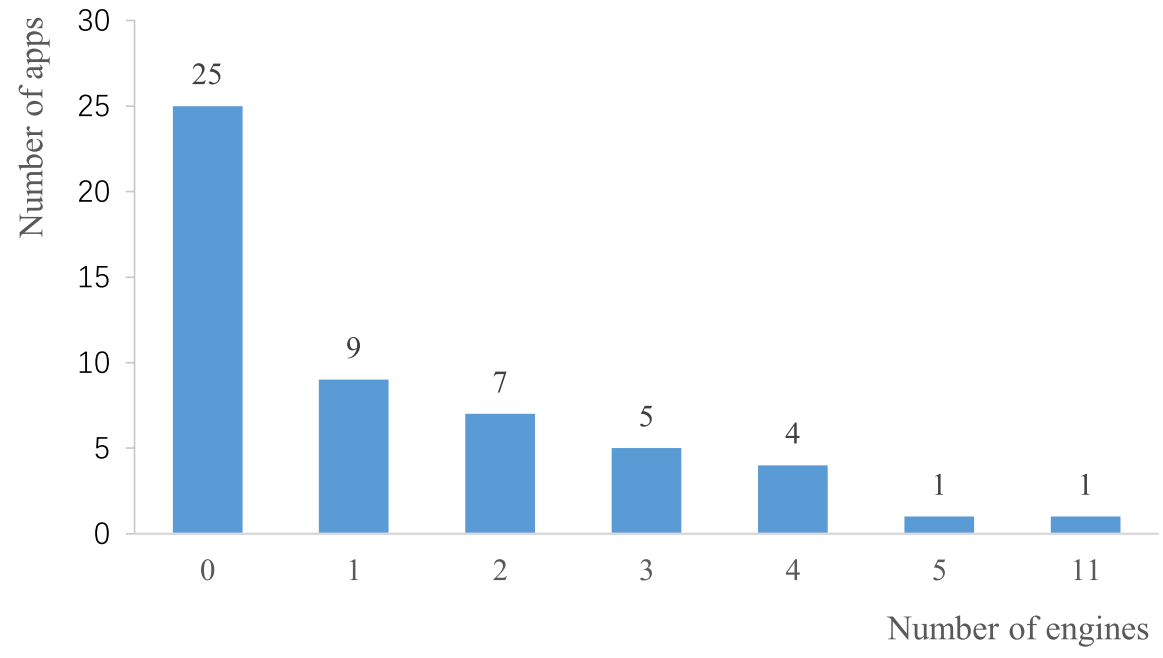

**Fig. 7** VirusTotal detection results for apps with negative red packet reviews. The x-axis indicates the number of anti-virus engines that flag apps as malicious, and the y-axis represents the number of the corresponding apps

#### Limitations

To the best of our knowledge, this work is the first attempt in the research community to investigate red packet fraud through user reviews. However, our work still has several limitations.

First, the user reviews collected in this study were solely sourced from Android apps, without considering apps from other operating systems such as iOS. This choice was made because the Android platform holds the largest market share worldwide. In future research, we plan to incorporate apps and user reviews from a broader range of platforms. Second, to identify user reviews related to red packets, we attempted to filter all collected reviews using keywords such as “red packet", “cash", “coins". However, this may have overlooked many reviews that implicitly express dissatisfaction with red packets without explicitly mentioning them, potentially leading to an underestimation of red packet-related issues. To mitigate this limitation, we extended the keyword set to include over 20 terms in both Chinese and English, such as “rewards", “points", “earn", “fake", “misleading", and “scam". Third, to identify negative red packet reviews, we manually labeled 2,500 negative reviews and 2,500 non-negative reviews as a dataset for training or fine-tuning sentiment classification models. The manual labeling process is inherently subjective, especially in the context of user-generated content where expressions of sentiment can be nuanced or ambiguous. As a result, inconsistencies and labeling errors may occur when different annotators interpret the same user review differently. To mitigate this issue, each review was labeled independently by two annotators. Only the reviews with consistent labels were included in our dataset. For instances where the labels differed, a discussion was conducted to reach a consensus, ensuring higher labeling reliability and consistency.

## Related work

### Fraudulent behaviors in mobile apps

In the mobile app ecosystem, various forms of fraud related to apps have been extensively explored, including ad fraud (Liu et al. 2014; Cao et al. 2021; Dong et al. 2018; Chen et al. 2019; Liu et al. 2020), fake apps (Hu et al. 2020; Tang et al. 2019; Malisa et al. 2016; Karunanayake et al. 2022) and other fraudulent behaviors (Cheng et al. 2025; Liu et al. 2016, 2019; Hu et al. 2021b; Martens and Maalej 2019; Chen et al. 2024).

For ad fraud, Liu et al. (2014) investigated ad placement fraud in mobile apps and proposed DECAF to detect five types of ad fraud (e.g., small ads, hidden ads, and intrusive ads, etc.) for Windows-based mobile platforms by analyzing visual UI layouts of app pages. Dong et al. (2018) revealed ad dynamic interaction fraud involving multiple UI states and proposed FraudDroid to detect such ad fraud as well as ad placement fraud in Android apps. For fake apps, Hu et al. (2020) explored the presence of squatting attacks in the mobile app ecosystem and proposed AppCrazy tool to identify fake apps with identifiers (e.g., app name or package name) that are confusingly similar to those of popular apps or well-known Internet brands. Karunanayake et al. (2022) proposed to apply the recent advances in deep learning methods to the problem of app counterfeit detection. They employed an approach that combines content embeddings and style embeddings, identifying 2,040 potential counterfeits that contain malware in a set of 49,608 apps that showed high similarity to one of the top-10,000 popular apps in Google Play Store. In addition to the aforementioned types of fraud, researchers have also uncovered other forms of fraudulent behaviors. For example, Hu et al. (2021b) systematically investigated the ecosystem of fraudulent dating apps and exposed the purpose of these apps, which lure users into purchasing premium/VIP services to start conversations with other fake female accounts in apps. Martens and Maalej (2019) studied the review fraud in mobile apps and developed a supervised classifier to automatically detect fake reviews in app stores. Cheng et al. (2025) conducted a study on red packet fraud in mobile apps and proposed a automated approach, ReckDroid, for detecting the red packet fraud.

Although red packets have been widely used in our daily apps, research on red packet fraud is still at an early stage in our research community. Cheng et al. (2025) conducted the first investigation into the phenomenon of red packet fraud and summarized three types of fraudulent behaviors, including aggressive advertising, malicious app, and malicious redirect. They primarily focus on the use of fake red packets as bait to deceive users and spread malicious content to them, whereas we investigate red packet fraud issues reported by users in user reviews, with a particular emphasis on fraudulent practices related to the amounts of red packets. Our research serves as a significant contribution to the further advancement and refinement of the field pertaining to red packet fraud.

### User review analysis

A number of studies have investigated and analyzed user reviews for mobile apps (Liu et al. 2023; Haggag et al. 2022; Li et al. 2020; Hu et al. 2021a; Aljedaani et al. 2022; Zhang et al. 2025; Le et al. 2023; Wang et al. 2020), which play a crucial role in the evolution of the mobile app ecosystem. Some of these studies analyze user reviews to propose solutions for improving the functionality or user interface of mobile apps. For examples, Li et al. (2020) analyzed bug-related user reviews of mobile apps and leveraged natural language processing techniques to extract valuable information from relevant user reviews for automated bug reproduction. Aljedaani et al. (2022) employed supervised learning techniques to automatically classify user reviews related to accessibility, thereby providing guidance for app developers in the development of building mobile apps with better accessibility.

Some studies uncover the issues related to security in apps through the analysis of user reviews, such as privacy leakage, privilege escalation, and app fraud. For example, Zhang et al. (2025) trained supervised classifiers to automatically identify privacy reviews from user reviews and designed an interpretable topic mining algorithm to effectively mining privacy concern topics contained in the privacy reviews. To understand why an app requests a permission, Wang et al. (2020) proposed a framework, called SmartPI, to automatically identify functionality-relevant user reviews and infer the permission implication of them, bridging the gap between the functionalities and the actual behaviors of an app. To uncover the app behaviors that violate market policies, Hu et al. (2021a) adopted text mining and natural language processing techniques to extract semantic rules from user reviews through a semi-automated process and analyzed 26 types of undesired behaviors that violate market policies, including security-related user concerns.

Although user reviews have been extensively studied from various perspectives, none of the existing works correlate user reviews to red packet fraud and none of them can be easily adopted or extended to investigate this issue.

## Conclusions

In this paper, we present the first analysis of user reviews related to red packets utilizing natural language processing and machine learning techniques, uncovering six categories of red packet fraud issues in Android apps from the user perspective. We first propose a novel automated approach, ReckDetector, to effectively identify and collect apps with red packets from popular Android app markets. Experiments on the dataset of hundreds of real-world apps have demonstrated that ReckDetector achieves higher performance than the state-of-the-art tool ReckDroid in identifying red packets. Then, we collect over 360,000 real user reviews for 334 apps with red packets from Google Play and three popular alternative Android app markets. Through the extraction and analysis of user reviews related to red packets, we found that red packet fraud is highly prevalent, significantly impacting user experience and damaging the reputation of apps. Moreover, red packets have been widely exploited by unscrupulous app developers as a deceptive incentive mechanism to entice users into completing their designated tasks, thereby maximizing their profits. Finally, we recommend that market regulators pay closer attention to red packet fraud issues reported by users in their reviews and strengthen the management of apps in mobile app markets.

**Acknowledgements**
We sincerely thank all the authors of the open-source tools ReckDroid, Droidbot, and google-play-scraper on GitHub. And we also thank "anonymous" reviewers for their insights.

**Author contributions**
All authors contributed to this manuscript and approved its submission. Yu Cheng participated in all the research work and drafted the manuscript. Prof. Xiaofang Qi made a decisive contribution to the content of research and provided critical revisions to the manuscript. Prof. Yanhui Li made substantial contributions to the technical methodology, research design, and manuscript revision.

**Funding**
This work is supported by the National Natural Science Foundation of China under Grant Nos. 61972082 and 62172202.

**Data availability**
We provide all datasets and source code used to conduct this study at https://github.com/AppFraud/ReckDetector.

## Declarations

**Competing interests**
We declare that we have no known competing financial interests or personal relationships that could have appeared to influence the work reported in this paper.